\documentclass[letterpaper]{article} 
\usepackage{aaai2026}  
\usepackage{times}  
\usepackage{helvet}  
\usepackage{courier}  
\usepackage[hyphens]{url}  
\usepackage{graphicx} 
\usepackage{natbib}  
\usepackage{caption} 
\usepackage{algorithm}
\usepackage{algorithmic}

\usepackage{amsmath}
\usepackage{amssymb}
\usepackage{mathtools}
\usepackage{amsthm}
\usepackage{dsfont}
\usepackage{xcolor}

\usepackage{csquotes}
\usepackage{newfloat}
\usepackage{listings}
\DeclareCaptionStyle{ruled}{labelfont=normalfont,labelsep=colon,strut=off} 
\floatstyle{ruled}
\newfloat{listing}{tb}{lst}{}
\floatname{listing}{Listing}
\theoremstyle{plain}

\theoremstyle{definition}

\theoremstyle{remark}

\newcommand\ourmethod{{\textrm{AutoSchA}}}
\newcommand\ourlayermethod{{\textrm{node isolation}}}

\newcommand{\Adj}{\mathbf{A}}

\newcommand{\G}{\mathcal{G}}
\newcommand{\V}{\mathcal{V}}
\newcommand{\E}{\mathcal{E}}

\title{AutoSchA: Automatic Hierarchical Music Representations \\via Multi-Relational Node Isolation}

\author {
    Stephen Ni-Hahn\textsuperscript{\footnote{Equal Contribution}\rm 1},
    Rico Zhu\textsuperscript{$^*$\rm 1},
    Jerry Yin\textsuperscript{\rm 2},
    Yue Jiang\textsuperscript{\rm 1},
    Cynthia Rudin\textsuperscript{\rm 1},
    Simon Mak\textsuperscript{\rm 1}
}
\affiliations {
    
    \textsuperscript{\rm 1}Duke University\\
    \textsuperscript{\rm 2}Stanford University\\
    \{stephen.hahn, rico.zhu\}@duke.edu
}

\usepackage{bibentry}

\begin{document}

\maketitle

\begin{abstract}
Hierarchical representations provide powerful and principled approaches for analyzing many musical genres. Such representations have been broadly studied in music theory, for instance via Schenkerian analysis (SchA). Hierarchical music analyses, however, are highly cost-intensive; the analysis of a single piece of music requires a great deal of time and effort from trained experts. The representation of hierarchical analyses in a computer-readable format is a further challenge. Given recent developments in hierarchical deep learning and increasing quantities of computer-readable data, there is great promise in extending such work for an automatic hierarchical representation framework. This paper thus introduces a novel approach, \ourmethod{}, which extends recent developments in graph neural networks (GNNs) for hierarchical music analysis. \ourmethod{} features three key contributions: 1) a new graph learning framework for hierarchical music representation, 2) a new graph pooling mechanism based on node isolation that directly optimizes learned pooling assignments, and 3) a state-of-the-art architecture that integrates such developments for automatic hierarchical music analysis. We show, in a suite of experiments, that \ourmethod{} performs comparably to human experts when analyzing Baroque fugue subjects.
\end{abstract}

\begin{links}
    \link{Supplemental Material}{github.com/stephenHahn88/AutoSchA_Supplement.git}
\end{links}

\section{Introduction}
\label{sec:intro_and_related_work}

Music theory is an art that aims to understand how music of various styles is designed, constructed, and composed. Whether it is understood implicitly or explicitly, one's own theory of music is vital for any musician to generate original works of art with influence from what they have heard in the past. Hierarchical structures provide an elegant framework for music representation and are broadly used in the music community. In particular, Schenkerian analysis (SchA) is a widely taught representation which fuses elements of melody, harmony, form, and counterpoint. Many composers use SchA to inform their compositions by enhancing their understanding of other works, incorporating structural ideas and techniques into their own \cite{schenker2000art, jackson2001heinrich}. While SchA was originally designed for Western common practice tonal music, aspects have been used in a broad range of musical genres from rock and jazz to Chinese opera and African folk music \cite{nobile2014, stock1993application, larson2009analyzing}.

In this work, we use machine learning to identify the hierarchical structure of a musical piece -- a task we call \textit{automatic hierarchical music representation}. Such automatic representations provide users a better understanding of the direction a piece can take, in turn leading to more convincing generation at larger scales. While previous work on machine models for music theory and generation have shown promise for small-scale structures and surface-level counterpoint \cite{ferreira2023generating, huang2018music, huang2019counterpoint, wu2019hierarchical}, there is scant literature on our target problem of automatic hierarchical music representation. This may be due to two inherent difficulties. First, to learn complex hierarchical structure, the employed model requires careful analyses from trained experts. Such training data are \textit{cost-intensive} to generate: an analysis of a single musical phrase can take longer than 30 minutes depending on complexity. Second, compactly representing hierarchical analyses in a computer-readable data format for model training remains a challenge, with many published analyses being incomplete and not showing all details needed to present the entire analysis in a computer-readable format.

There has been recent promising work on hierarchical deep learning aimed at addressing these difficulties. In particular, graph neural networks (GNNs) have made strides in hierarchical data processing with pooling methods such as Diffpool \cite{diffpool-18} and top-$k$ pooling \cite{topk-18}. Given recent developments in hierarchical deep learning and the increasing amount of computer-readable data, there is promise in extending such work for automatic hierarchical representation of music. 

We thus introduce the first deep learning model for SchA, called \textit{\ourmethod{}}, which extends recent developments in GNNs for effective automatic music representation. Our novel contributions include 1) a new graph learning framework for hierarchical music representation, 2) a new graph pooling mechanism based on node isolation that directly optimizes learned pooling assignments, and 3) a state-of-the-art architecture integrating such developments for automatic hierarchical music analysis.

Section \ref{sec:related_work} briefly describes SchA, previous computational models, and existing models for graph pooling. Section \ref{sec:methodology} presents our methodology, including the representation of SchA as a graph pooling problem, and the components of our novel modeling architecture. Section \ref{sec:experiments} describes our experiments, which show that we perform comparably to humans in analyzing Baroque fugue subjects. This includes ablation experiments that investigate the most important features for computational SchA and compare with other baseline deep learning models. Our results show great promise in the future of AI-assisted SchA.

\section{Related Work}
\label{sec:related_work}

\subsection{Hierarchical Music Analysis}
\label{subsec:scha_background}

Hierarchical music representations are broadly used in many fields of music theory. For instance, \citet{hepokoski2006elements, caplin2013analyzing} focus on form or thematic structure, \citet{rothstein1989phrase, lerdahl1996generative, foscarin2023predictingmusichierarchiesgraphbased} focus on rhythmic, phrase, or grammar-based structure, and \citet{schenker2001free, cadwallader1998analysis} focus on the linear/harmonic structure. In what follows, we focus on SchA, a hierarchical linear-harmonic framework first codified by Heinrich Schenker aiming to capture how the music analyst hears the progression of musical harmonies on various levels of structure as they are ``unfolded'' by linear melodic motions. 

\begin{figure}[]
    \centering
    \includegraphics[width=0.45\textwidth]{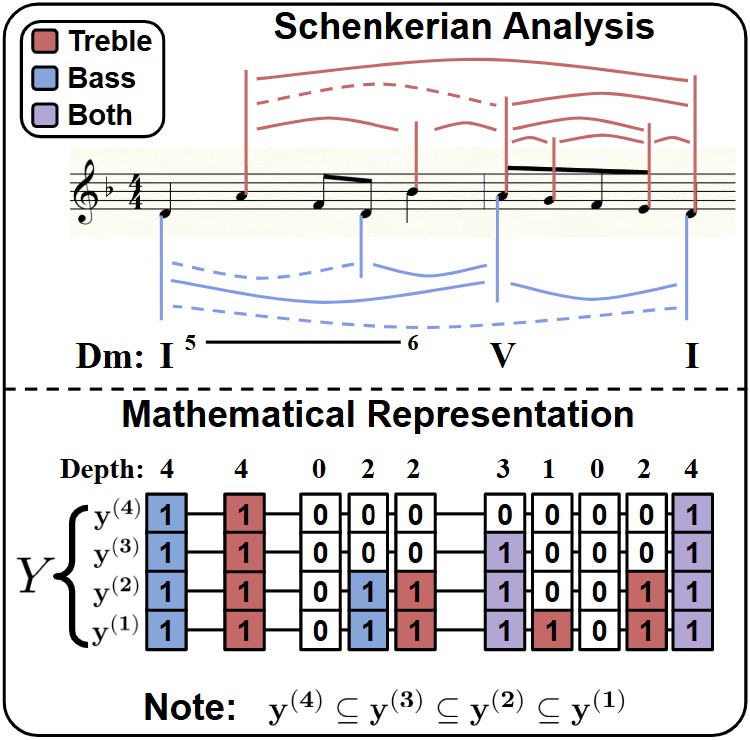}
    \vspace{-2mm}
    \caption{Top: SchA, Pachelbel's \textit{Primi Toni} No. 1, fugue subject. Bottom: Mathematical representation of SchA as several bit arrays denoting whether notes belong in a certain depth. Note that higher depths are included in lower depths (the first note is included in all depths from 0-4).}
    \label{fig:scha_example}
    \vspace{-3mm}
\end{figure}

Figure \ref{fig:scha_example} displays an example SchA for a Pachelbel fugue subject, outlining the hierarchy of tones and the harmonies these tones produce over time. Longer note stems indicate deeper levels of structure, slurs indicate prolongations between notes at a given level, and dotted slurs indicate prolongations between notes of the same pitch. Roman numerals describe harmonies unfolded by melodic tones. The analysis shows that the first four notes outline the root position tonic (``I'') triad, the fifth note destabilizes the tonic function with a 5--6 exchange as it moves towards the dominant ``V,'' which is expressed through the subsequent descending four notes. The final note shows a return to ``I.'' The downward and upward note stems indicate the bass and treble voices respectively, with the bass voice tending to outline harmonies with larger leaps and the treble and inner voices tending to act with smoother linear motions. In our example the bass follows the roots of the I-V-I harmonies (D-A-D) and the treble moves entirely stepwise (A-B$\flat$-A-G-F-E-D).

SchA depends on the various structural levels (``depths'') of the music. The music as written in the score is the ``foreground,'' which hosts all details that make a piece unique. As layers of decoration and relatively less structural tones are removed, deeper levels of ``middleground'' structure are revealed until finally we reach the deepest level, the ``background'' or \textit{Ursatz} structure, often represented with open note heads. For our model, notes on the foreground are considered depth 0, notes that have a stem are considered part of depth 1, notes with relatively longer stems are included in depth 2, etc., until the deepest level is reached. In Figure \ref{fig:scha_example}, the maximal depths of each note are notated at the bottom. This depth notation is not typically explicit in SchA. The pre-penultimate note (F4) belongs to depth 0 in the treble voice, but it does not belong in depth 1. The fourth-to-last note (G4) belongs to depths 0 and 1, but not depth 2, since its analytical stem is at the shortest level. Corresponding notes in the bass and treble such as the first two notes share the same maximal depth.

\begin{figure*}[t]
    \centering
    \includegraphics[width=1.0\linewidth]{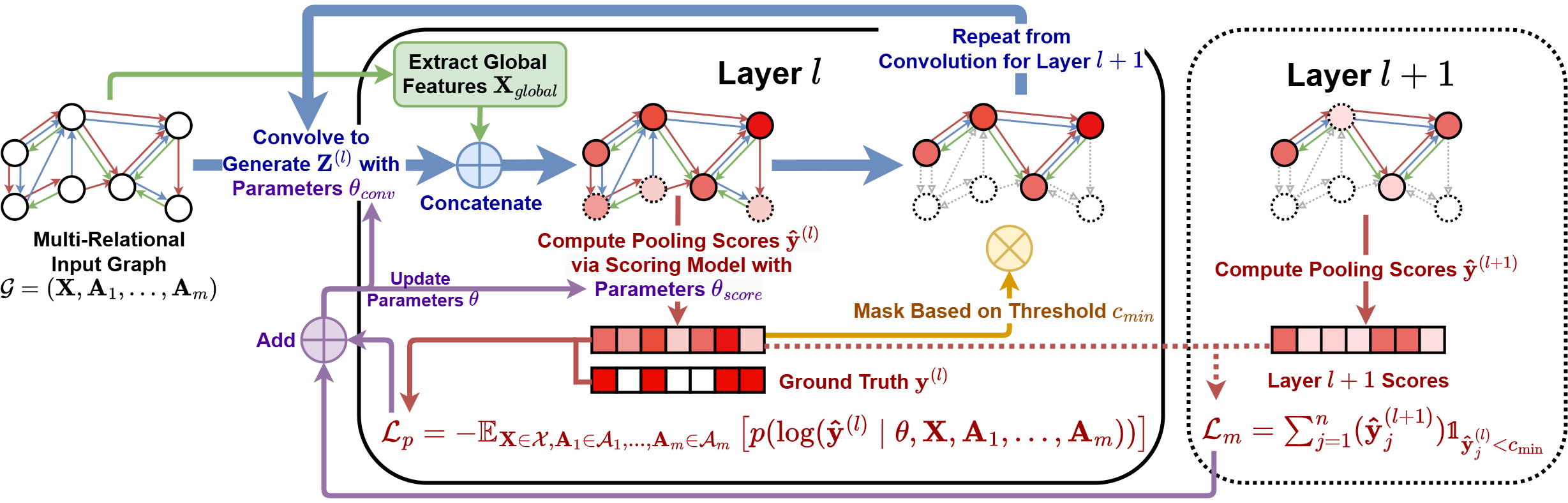}
    \caption{Overview of model architecture: diagram for the pooling GNN. Blue arrows indicate the major flow of GNN convolution, green arrows represents global feature extraction, yellow indicates threshold masking, red describes information regarding the scoring model and loss functions, and purple arrows show the flow of backpropogation.  Given input graph $\G = (\mathbf{X}, \Adj_1, ..., \Adj_m)$, we first perform a directed multi-relational convolution (parameterized by $\theta_{conv}$) over input features $\mathbf{X}$ to generate node embeddings $\mathbf{Z}^{(l)}$, and global feature matrix $\mathbf{X}_{\text{global}}$ (see Figure \ref{fig:global-features} for details). $\mathbf{Z}^{(l)}$ and $\mathbf{X}_{\text{global}}$ are concatenated and passed to the scoring model (parameterized by $\theta_{score}$), generating pooling scores $\mathbf{\hat{y}}^{(l)}$. We aim to minimize the cross-entropy $\mathcal{L}_p$ between the pooling scores and ground truth assignments. Based on $\mathbf{\hat{y}}^{(l)}$ and threshold $c_{min}$, the nodes are masked or ``isolated'' for the next GNN layer $l+1$, to which $\mathbf{Z}^{(l)}$ is passed in place of $\mathbf{X}$. To ensure monotonicity in pooling scores, we add a regularizer term $\mathcal{L}_m$ computed from scores at layer $l$ and layer $l+1$.}
    \label{fig:pooling-gnn}
    \vspace{-3mm}
\end{figure*}

\subsection{Computational Schenkerian Analysis}
\label{subsec:computational_scha}

While structural analyses such as SchA provide vital insights into symbolic music, there is a paucity of research on computational SchA. The vast majority of past approaches rely on heuristics or rule-based systems due to the absence of computer-readable data \cite{kassler1975proving, frankel1976lisp, smoliar1979computer, mavromatis2004parsing, gilbert2007probabilistic}. The first steps toward a data-driven approach were taken by \citet{marsden2010schenkerian}, who introduced a ``goodness metric'' based on a corpus of six Mozart analyses. This metric was used to determine high scoring analyses out of a prohibitively large search space of possible analyses. 

Most notably, Kirlin introduced a probabilistic approach based on a new corpus of 41 SchAs \cite{kirlin2011probabilistic, kirlin2014data, kirlin2015learning}. Such models employ random forests to predict how likely certain notes \textit{prolong} others, where a prolongation is defined as two notes $\text{note}_1$ and $\text{note}_\alpha$ that are more structural than a group of intervening notes $(\text{note}_2,..., \text{note}_{\alpha-1})$. These analyses were represented as restricted versions of Yust's maximal outerplanar graphs (MOPs; \citealp{yust2015voice}). 

While their work is based in Schenkerian theory, the model of \citet{kirlin2015learning} does not perform a proper Schenkerian analysis, nor do the authors claim that it does so. Their model reads a melodic line as a single theoretical voice, often misinterpreting the compound polyphonic nature of the music. In fact, it is incapable of reading multiple musical lines simultaneously, which is vital to SchA even in monophonic settings where music is notated as a single voice. Without an understanding of compound melody and counterpoint between voices, it is impossible to describe musical structure as we hear it.

\subsection{Graph Neural Networks for Graph Pooling}
\label{subsec:gnn_for_graph_pooling}

We introduce a model for computational SchA as a graph pooling problem framed using GNNs. GNNs learn mappings from a node parameter space to some latent space, where the geometric relations of the embedded nodes reflect the structure of the original graph \cite{grl-2017}. In graph classification, it is necessary to condense node-level representations into graph-level representations, an operation termed \textit{graph pooling}. Early approaches mainly focused on adapting existing graph clustering algorithms \cite{bruna-13, defferrard-2017}, e.g., the Graclus algorithm \cite{graclus-07}. The current state-of-the-art in graph pooling operators permit dynamic learning of optimal pooling assignments to adapt to downstream tasks \cite{pooling-survey-21}. Of this new family of learnable pooling operators, two main approaches have emerged: clustering approaches, which learn grouping matrices from the node embedding space \cite{diffpool-18, mincutpool-20, bacciu-19, structpool-20, lapool-19, dmon-23}, and top-$k$ approaches, which fit a scoring function over the node set and prune the lowest scoring nodes, the number of which to prune at each layer being determined by a predefined percentage \cite{graphunets-19, topk-18, sagpool-19, asap-20, hgpsl-20} or score threshold \cite{min-score-pool-19}.

Real-world networks often involve modeling the heterogeneous interactions between a collection of entities, motivating graph learning algorithms that operate on a multi-relational domain. \citet{rgcn-17} and \citet{multigcn-19} propose methods to compute node embeddings that aggregate information across all relations present in a heterogeneous graph. \textit{Pooling} multi-relational graphs, however, is relatively underexplored, and existing methods cannot be applied directly for SchA. For instance, \citet{user-hetergpool-21} describes pooling for a heterogeneous vertex set, not edge sets as in SchA. \citet{multigcn-19} first compute a consensus graph over all edge types and then perform a manifold ranking \cite{manifold-ranking-2003} to compute node saliency, but this approach is structural and not learnable in that it does not incorporate node embedding information to compute the pooling filtration. Finally, \citet{multiview-gpool-21} proposes an algorithm that is able to leverage node information, but is non-adaptive in that the number of clusters the model learns is fixed as a hyperparameter.

\section{Methodology}
\label{sec:methodology}



\subsection{Notation and Problem Formulation}
\label{subsec:problem_formulation}

We consider multi-relational graphs of the form $\G = (\V, \E, \mathbf{X})$, where $\V$ is the set of $n$ vertices $\V = \{v_1, ..., v_n\}$ indexed by $j$, $\E = \{E_{1}, ..., E_{m}\}$ is the union of edge sets over all $m$ edge types (indexed by $i$), and $\mathbf{X} \in \mathbb{R}^{n \times p}$ is the input feature matrix where the $j$th row describes features of the corresponding $j$th node. The number of edge types is constant throughout the entire dataset; thus, any given graph $\G$ can be fully characterized by $\mathbf{X} \in \mathbb{R}^{n \times p}$, as well as adjacency matrices $\Adj_{1}, ..., \Adj_{m} \in \mathbb{R}^{n \times n}$. 

We express a corresponding Schenkerian analysis of depth $d$ as a sequence of bit arrays $Y = (\mathbf{y}^{(1)}, ..., \mathbf{y}^{(d)})$, each indicating whether particular vertices belong within a particular depth (see Figure \ref{fig:scha_example}). Higher depths are more sparse, containing more structural notes. Since Schenkerian analysis can be understood as a recursive process where all notes that are removed at a given level cannot appear in deeper levels, the sequence is nested: $\mathbf{y}^{(l)} \subseteq \mathbf{y}^{(k)}$ for all $l \geq k$.

\subsection{Graph Representation for Symbolic Music}
\begin{figure}
    \centering
    \includegraphics[width=\linewidth]{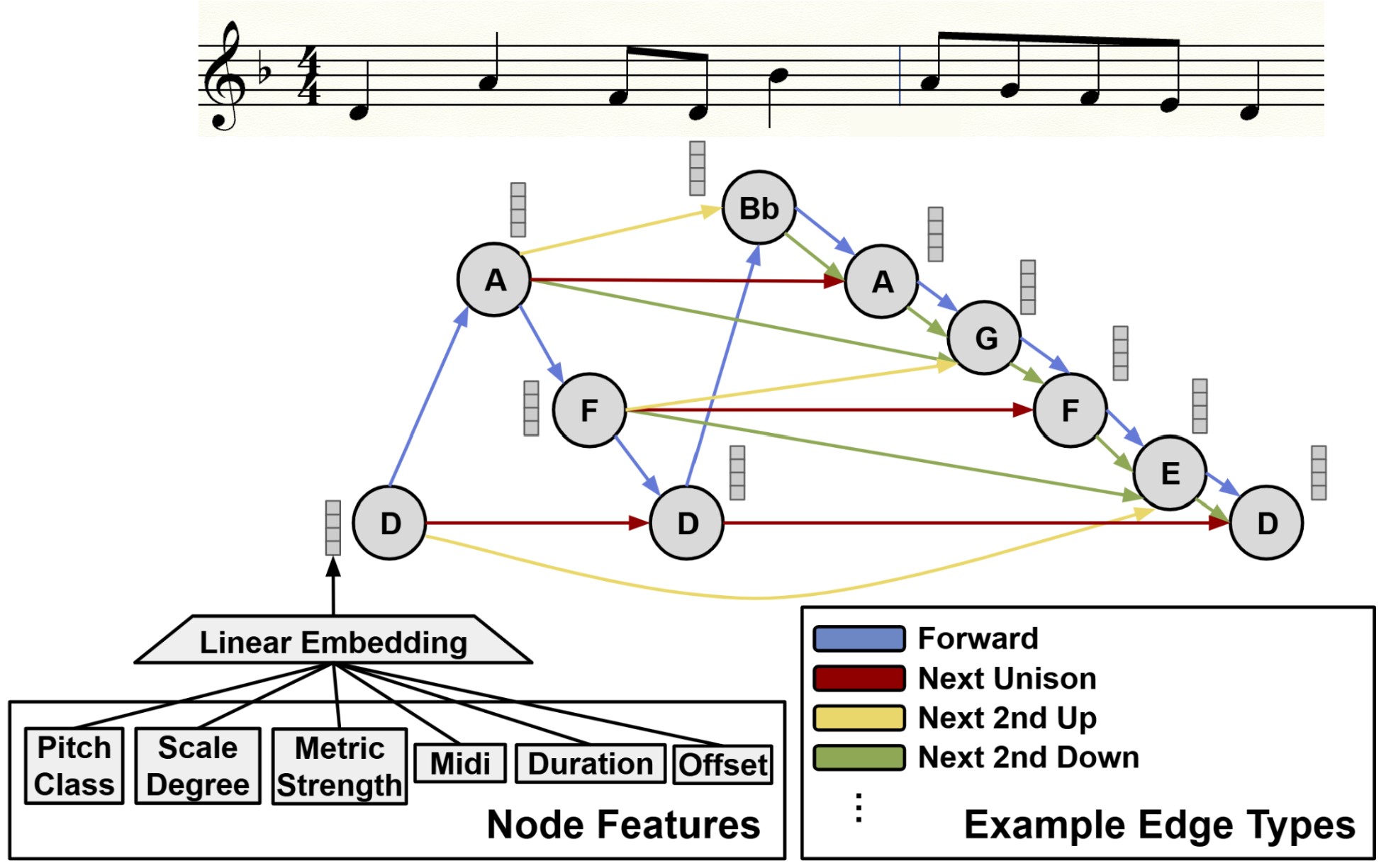}
    \caption{\textit{Primi Toni} No. 1 as a multi-relational graph.}
    \label{fig:music_to_graph}
    \vspace{-3mm}
\end{figure}

In order to process music through our GNN, we first convert musical scores to multi-relational graph data structures (see Figure \ref{fig:music_to_graph}). Following \citet{jeong2019graph}, we represent nodes as musical notes and edges as relationships between notes; going forward, ``note'' and ``node'' are used interchangeably.

In our representation, nodes are described by various continuous and categorical features. We include three pitch features: 1) pitch class: the discrete note pitch name (e.g., A$\sharp$ vs. D$\flat$, etc.); 2) normalized midi pitch, which encodes a proxy for a note's location on the piano; and 3) scale degree: the note's relationship to the home key. We also include three rhythmic features: 1) normalized duration: the relative length of a note to other notes in the excerpt; 2) normalized offset: the relative location in time of a note; and 3) metric strength: the relative strength of the onset of a note (e.g., downbeats vs. offbeats).

For edge types, we first adopt the edges described by \citet{jeong2019graph}. These include 1) onset edges, which relate notes that begin at the same time; 2) forward edges, which relate notes with notes that immediately follow in time; 3) voice edges, which are similar to forward edges, but for notes within the same contrapuntal line; 4) rest edges, which are similar to forward edges, but connecting notes separated by rests; 5) sustain edges, which relate all notes that occur while one is held; and 6) slur edges, which relate notes that share a slur. Because music only flows forwards in time, the forward edge graph is acyclic.

To address deeper structural connections, we add novel \textit{intervalic} edges. For each note, we create an edge to the next note up or down a certain diatonic interval if it exists. For example, in Figure \ref{fig:music_to_graph}, different edge types describe the relationships between the notes of the score. The first note (D4) is connected with the penultimate note (E4) by a ``next 2nd up'' edge. Similarly, the third note (F4) is connected to the 7th note (G4) with the same edge and to the penultimate note (E4) with a ``next 2nd down'' edge.

\subsection{Pooling Structure} 
\label{subsec:pooling_structure}

One way to interpret SchA is to ``remove'' structurally less important notes at various depths, only keeping those describing linear and harmonic function of the music at each depth. Top-$k$ pooling approaches provide a natural formulation for this task as they follow a related procedure of node removal. However, such methods have not been used for heterogeneous graphs and often require $k$ (the percentage of nodes to drop at each depth) to be set in advance as a hyperparameter. For SchA, this $k$ can vary from piece to piece and from depth to depth, necessitating an adaptive method for removing nodes.

\begin{figure*}[t]
    \centering
    \includegraphics[width=1.0\linewidth]{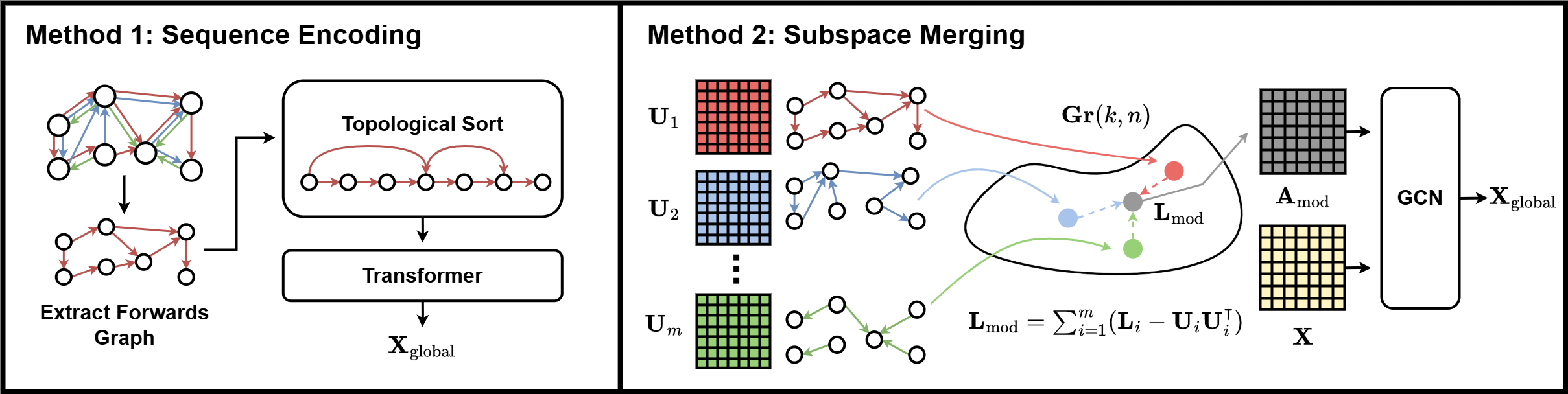}
    \caption{Two approaches to compute the global embedding, $\mathbf{X}_{\text{global}}$. (1) A sequential approach, leveraging a transformer to encode long-range dependencies. We obtain a canonical sequence representation of our graph via the topological order of the forwards edges. (2) A subspace merging approach, based on computing a unified global topology fusing all edge types. We compute a fused global graph $\mathbf{A}_{\text{mod}}$ given fixed embeddings $\mathbf{U}_1, ..., \mathbf{U}_m$ from the Laplacians $\mathbf{L}_1, ..., \mathbf{L}_m$, and then convolve the original features $\mathbf{X}$ over $\mathbf{A}_{\text{mod}}$. Further details are in the technical appendix.}
    \label{fig:global-features}
    \vspace{-4mm}
\end{figure*}

While pooling is typically a byproduct in classification tasks, it is vital to SchA. To our knowledge, our model is the first to directly optimize a sequence of pooling assignments as its main objective. This section introduces our novel multi-relational pooling layer based on \ourlayermethod{}, a general framework that can explicitly learn pooled node representations for classification based on any given criteria for SchA or related tasks.

\subsubsection{Multi-Relational Graph Convolution}
\label{subsubsec:multi-relational_convolution}

The first step in processing node embeddings for SchA is graph convolution (recall Figure \ref{fig:pooling-gnn}). The standard graph convolution operator introduced in \citet{kipf-welling-17} is given by
\begin{equation*}
    \begin{split}
        \mathbf{Z}^{(l+1)} &= \sigma(\mathbf{\tilde{D}}^{-\frac{1}{2}}\mathbf{\tilde{A}}\mathbf{\tilde{D}}^{-\frac{1}{2}}\mathbf{Z}^{(l)}\mathbf{W}^{(l+1)}),
    \end{split}
\end{equation*}
where $\mathbf{Z}^{(l)}$ is the node embedding matrix after $l$ GNN layers ($\mathbf{Z}^{(0)}=\mathbf{X}$), $\mathbf{\tilde{A}} = \mathbf{A} + \mathbf{I}_n$ is the adjacency matrix with self-loops, $\mathbf{\tilde{D}}$ is the associated degree matrix of $\mathbf{\tilde{A}}$, $\mathbf{W}^{(l)} \in \mathbb{R}^{p\times q}$ is a matrix of learnable parameters for layer $l$ where $p$ and $q$ are hyperparameters, and $\sigma$ is a non-linear activation function (e.g., ReLU). The expression $\mathbf{\tilde{D}}^{-\frac{1}{2}}\mathbf{\tilde{A}}\mathbf{\tilde{D}}^{-\frac{1}{2}}\mathbf{Z}^{(l)}$ spreads node features to their neighbors while scaling them based on node degree. This allows each node to learn from its neighbors without letting high-degree nodes overwhelm the information. The result is then linearly transformed by $\mathbf{W}^{(l+1)}$ to update the node embeddings.

A natural extension of this operator to the multi-relational setting is given in \citet{rgcn-17}. The matrix form of the node embedding equation is
\begin{equation*}
    \begin{split}
        \mathbf{Z}^{(l+1)} &= \sigma\left(\sum_{i=1}^m\mathbf{\tilde{D}}_i^{-\frac{1}{2}}\mathbf{\tilde{A}}_i\mathbf{\tilde{D}}_i^{-\frac{1}{2}}\mathbf{Z}^{(l)}\mathbf{W}^{(l+1)}_{i}\right).
    \end{split}
\end{equation*}
More generally, the sum over edge types can be replaced with any aggregation method (e.g., concatenation). We extend the multi-relational convolution operator to the directed case, letting $\mathbf{\tilde{M}} = \mathbf{\tilde{D}}^{-\frac{1}{2}}\mathbf{\tilde{A}}\mathbf{\tilde{D}}^{-\frac{1}{2}}$ denote the normalized forward edge adjacency matrix as in \citet{dir-gcn-23}:
\begin{equation*}\label{eq:dir-rgcn}
    \begin{split}
        \mathbf{Z}^{(l+1)} \hspace{-1mm}&= \sigma\hspace{-1mm}\left(\sum_{i=1}^m \left(\alpha\mathbf{\tilde{M}}_{i}\mathbf{Z}^{(l)}\mathbf{W}^{(l+1)}_{i, \rightarrow} + \bar{\alpha}\mathbf{\tilde{M}}_{i}^\intercal\mathbf{Z}^{(l)}\mathbf{W}^{(l+1)}_{i, \leftarrow} \right)\hspace{-1mm}\right)\hspace{-1mm}.
    \end{split}
\end{equation*}
Here $\mathbf{W}_{i, \rightarrow}, \mathbf{W}_{i, \leftarrow}$ denote the learnable weights for the forward and backward direction edges for edge type $i$, and $\alpha \in [0, 1]$ is a hyperparameter weighing the forwards and backwards edges in tandem with $\bar{\alpha} = 1 - \alpha$.

\subsubsection{Node Isolation for Multi-Relational Pooling}
\label{subsubsec:multi-relational_pooling}
One of the main challenges in SchA is that the number of nodes we wish to drop at each level varies with respect to the input graph. For example, a passage with a high density of neighbor or repeated tones would result in a large percentage of notes being dropped, as neighbor tones are non-structural with regard to the foreground of the piece. On the other hand, a passage with a variety of harmonic shifts along with few non-harmonic tones (NHTs) and repetitions would drop a low percentage of notes at each level of Schenkerian depth.

We now outline a novel node downsampling mechanism, ``\ourlayermethod{}.'' We first compute a score $\mathbf{\hat{y}}_j$ for each node in the input graph \citep[similar to][]{sagpool-19} using the multi-relational convolution of \citet{dir-gcn-23} described earlier:
\begin{equation*}
    \begin{split}
        \mathbf{\hat{y}} &= \sigma\left(\sum_{i=1}^m \left(\alpha\mathbf{\tilde{M}}_{i}\mathbf{Z}\mathbf{W}_{i, \rightarrow} + \bar{\alpha}\mathbf{\tilde{M}}_{i}^\intercal \mathbf{Z}\mathbf{W}_{i, \leftarrow} \right)\right),\\
    \end{split}
\end{equation*}
where $\mathbf{W}_{i, \rightarrow},  \mathbf{W}_{i, \leftarrow} \in \mathbb{R}^{p\times 1}$ and $\sigma : \mathbb{R}^{n}\rightarrow [0, 1]^{n}$ represents the sigmoid function. Next, we prune the adjacency matrices $\mathbf{A}_1, ..., \mathbf{A}_m$ in accordance with the indicator vector $\mathbf{idx} = \mathds{1}_{\mathbf{\hat{y}}_j \geq c_{\min}}$ for some minimum threshold hyperparameter $c_{\min} \in (0, 1)$:
\begin{equation}
    \begin{split}
        \mathbf{A}'_i &= \left(\mathbf{A}_{i} \odot \mathbf{idx}\right)^\intercal \odot \mathbf{idx},\\
        \mathbf{Z}' &= \mathbf{Z} \odot \mathbf{\hat{y}},
    \end{split}
    \label{eq:scoring-gnn}
\end{equation}
where $\odot$ denotes a row-wise product. The indicator vector, $\mathbf{idx}$, masks the rows and columns of the adjacency matrix to ensure all edges to and from isolated nodes are pruned. The node embeddings are then weighted by their scores $\mathbf{\hat{y}}$ before being passed to the next convolutional layer.

While the nodes below the scoring threshold remain in the graph, by pruning edge connections, we create isolated nodes that cannot affect other nodes; no message passing occurs in or out of the isolated nodes. 

Through this approach, our method easily extends existing classification models with access to ground truth node values to learn pooling assignments. We train learnable weights $\mathbf{W}$ of our pooling assignment model to minimize the binary cross-entropy between the generated score vector $\mathbf{\hat{y}}$ and the underlying ground truth  $\mathbf{y}$ at each depth (see Figure \ref{fig:pooling-gnn}). The objective is then summed over all $d$ depths to obtain the pooling loss term:
\begin{equation*}
    \begin{split}
        \mathcal{L}_p(\theta) &= \hspace{-1mm}-\mathbb{E}_{\mathbf{X} \in \mathcal{X}, \mathbf{A}_{1,...,m} \in \mathcal{A}_{1,...,m}}\hspace{-1mm}\left[\log p(\mathbf{\hat{y}} \hspace{-1mm}\mid\hspace{-1mm} \theta, \mathbf{X}, \mathbf{A}_1, ..., \mathbf{A}_m) \right]\\
        &= -\sum_{l=1}^{d} \mathbf{y}^{(l)} \log (\mathbf{\hat{y}}^{(l)}) + (1-\mathbf{y}^{(l)}) \log (1 - \mathbf{\hat{y}}^{(l)}),
    \end{split}
\end{equation*}
where $\theta$ represents learnable parameters $\mathbf{W}^{(l)}_{i, o}$ for all $1\leq i\leq m$, $1\leq l \leq d$, and $o\in \{\leftarrow,\rightarrow\}$. Note that the variables of $\mathbf{\hat{y}}$ we want to optimize are contained in $\theta$. 

To ensure $\mathbf{\hat{y}}^{(1)}, \dots, \mathbf{\hat{y}}^{(l)},\dots, \mathbf{\hat{y}}^{(d)}$ is a valid filtration, we introduce a second novel loss term that acts as a monotonicity regularizer. This new loss term encourages nodes that have a score below the cutoff threshold at level $l$ to remain below the threshold at level $l+1$ (and thus at all remaining levels), 
which leads to the loss term:
\vspace{-2mm}
\begin{equation*}
    \begin{split}\label{eq:monotonicity-loss}
        \mathcal{L}_m(\theta) &= \sum_{l=2}^{d-1} \sum_{j=1}^n \mathbf{\hat{y}}_{j}^{(l+1)} \cdot\mathds{1}_{\mathbf{\hat{y}}_{j}^{(l)} < c_{\min}}.
    \end{split}
\end{equation*}

\noindent Thus, the complete objective is $\mathcal{L}(\theta) = \mathcal{L}_p(\theta) + \mathcal{L}_m(\theta)$.

\subsection{Global Feature Aggregation}
\label{subsec:global_feature_aggregation}

In addition to local information such as interval relations between nearby notes, global information (e.g., key, large-scale tonicization) is also vital to SchA. Thus, the model requires understanding of global feature input in order to create a suitable node scoring function. We investigate two approaches below, based on sequential and topological views of the music graph. Figure \ref{fig:global-features} visualizes the two proposed methods side by side.

\subsubsection{Sequential Approach}
\label{subsubsec:sequential_approach}

Previous approaches for algorithmic symbolic music composition and harmonization treat music as a sequence of discretized tokens. Transformer models in particular demonstrate strong ability to leverage long term harmonic information encoded within a passage globally in performing composition tasks \cite{huang2018music, von2022figaro}. 

We use a transformer encoder on the input features to a given layer $l$ (see Method 1 of Figure \ref{fig:global-features}). To compute an appropriate sequence representation of our graph, we leverage the fact that the forward edge graph should be a directed acyclic graph by first performing a topological sort of our nodes, breaking ties in vertical order (treble to bass):
\begin{equation*}
    \begin{split}
        \mathbf{x}^{(l)}_{\text{global}} &= \text{Transformer}\big(\text{Topological-Sort}(\mathbf{Z}^{(l)})\big),\\
        \mathbf{Z}'^{(l)} &= 
        \begin{bmatrix} \vphantom{\mathbf{x}^{(l)}_{\text{global}} \cdot \mathbf{y}^{(l-1)}_n} \mathbf{Z}^{(l)}_1 \\
            \vdots \\
            \mathbf{Z}^{(l)}_n \vphantom{\mathbf{x}^{(l)}_{\text{global}} \cdot \mathbf{y}^{(l-1)}_n}
        \end{bmatrix} \mathbin\Vert \begin{bmatrix}
             \mathbf{x}^{(l)}_{\text{global}} \cdot \mathbf{\hat{y}}^{(l-1)}_1 \\
             \vdots \\
             \mathbf{x}^{(l)}_{\text{global}} \cdot \mathbf{\hat{y}}^{(l-1)}_n \\
        \end{bmatrix},
    \end{split}
\end{equation*}
where $\mathbin\Vert$ represents column concatenation.

Here, $\text{Topological-Sort}(\cdot)$ performs a topological sort of the rows of $\mathbf{Z}^{(l)}$. $\mathbf{Z}'^{(l)}$ is then passed into the scoring GNN to compute pooling scores. We then reweigh the global embedding with the node attention scores from the previous layer. This prevents the global embeddings from disrupting monotonicity by reducing the influence of nodes that have already been dropped out or are close to dropping out. All together, this approach helps the model understand each note's place within the whole of the piece. 

\subsubsection{Subspace Merging Approach}
\label{subsubsec:subspace_merging_approach}

Alternatively, the subspace merging approach aims to find a unified global topology that can capture the node connections over all edge types (see Method 2 of Figure \ref{fig:global-features} and the technical appendix for full technical details). For simplicity, we focus on the undirected representation of our input graphs, $\hat{\Adj}_i$ (computed as the logical \textit{or} between $\Adj_i$ and $\Adj^\intercal_i$), which we use to generate a fused global graph $\Adj_{\text{mod}}$. Using this global topological information, we then convolve the initial embeddings over the fused graph, generating node-level features which integrate information over all layers of the graph. We find that this approach improves predictive accuracy over the baseline model with a comparable number of parameters.

\subsection{Voice Assignment and Inference}
\label{subsec:voice_assignment}

SchA makes the distinction between different theoretical voices. Western tonal music particularly relies on distinct structural roles for treble and bass voices. Inner voices may also host important motivic and structural support, but generally not at the same level as the indicated outer voices. We thus pose our task as a multi-label prediction task over all graph nodes, with class options in \{\textit{treble}, \textit{bass}, \textit{inner}\}. Since voice assignment is determined exclusively using information from the first GNN convolution layer, we need only to generate the voice assignment once and propagate the assignment to each depth. We trained a Directed R-GCN as described in Section \ref{subsec:pooling_structure} with sigmoid activation to perform this task.

We first predict the voice assignment for each node in the input graph and then compute scores for each musical depth represented in our model. For both voice and depth inference, a human expert may choose their preferred threshold $c_{\min}$ to produce their ideal results. If a higher number of depths are required or ambiguities arise in the highest depth given by the model, the relative confidence of scores can be used to inform expert decisions.

\section{Experiments and Results}
\label{sec:experiments}
\subsection{Experimental Design}

We present a suite of experiments to assess performance of our methodology. We use the Schenkerian analysis dataset introduced by \citet{nihahn2024newdatasetnotationsoftware}. The dataset consists of $>$140 analyses in computer-readable format. Particularly, we focus on 88 analyses of fugue subjects by Bach and Pachelbel as they are the most abundant and relatively homogeneous subset of analyses in the dataset. Because we are estimating structure for each note individually, this totals 3807 inferences per estimated depth (there are 3807 total notes, each note could be in each depth). To augment the data, we transpose the analyses to 12 musical keys, leading to a total 45,684 inferences per estimated depth. 

\subsubsection{Relative Performance}
\label{subsec:baselines}

To measure the relative performance of our model, we compare with other deep sequence and graph-based approaches. We separately train each baseline for each level of SchA to minimize the binary cross-entropy of node scores for specific depths. We compare with two non-graph-based models, a multi-layer perceptron (MLP) and transformer. We encode the nodes with positional information given by the topological order of the forwards graph as in Section \ref{subsec:global_feature_aggregation}. We also compare with three graph-based models: GCN \cite{kipf-welling-17}, Graph Attention Networks (GATs) \cite{gat-18}, and R-GCN \cite{rgcn-17}. For the GCN and GAT, we use the undirected representation of our graphs with adjacency matrix equal to the union of all edge types. Accuracy is measured by the proportion of node scores on the correct side of the threshold as compared to the ground truth of expert-annotated SchA, and monotonicity loss was defined previously in Section \ref{eq:monotonicity-loss}). We report the means and standard deviations of the maximum accuracy and lowest monotonicity loss attained across test samples in Table \ref{table:baselines_vs_model}.

\subsubsection{Human Experiments}
To assess the subjective performance of \ourmethod{}'s analyses, we surveyed five experts in music theory and Schenkerian analysis. The survey presented 9 scores and midi recordings, each with three corresponding Schenkerian analyses. We compare the human analysis from the dataset, \ourmethod{}'s analysis, and an analysis purposefully designed to be unmusical, which we denote as ``Flawed'' analyses. We do not compare against \citet{kirlin2015learning} because their model does not perform proper Schenkerian analysis (see Section \ref{subsec:computational_scha}). Analyses from \ourmethod{} were created by choosing an optimal threshold at each depth, chosen by an expert in Schenkerian analysis. It is important to note that this evaluation represents the largest number of Schenkerian experts ever polled for experiments in computational SchA (experiments by \citealt{kirlin2015learning} relied on 3 experts). Experts with graduate-level training in the relatively small field of SchA are extremely difficult to find, and in this case, they each provided hours of effort analyzing SchA attempts.

The three analyses for each score were presented in random order to avoid positional bias. For each analysis, we assessed overall quality as a letter grade, asking ``What letter grade would you assign this analysis'' as if grading a student (A+, A, A-, B+, etc.). To judge perceived musicality of an analysis we asked, ``How would you score the musicality of this analysis on a scale of 0 (not musical) to 10 (very musical)?''. We performed a Turing test, asking ``How certain are you that the analysis was written by a human on a scale of 0 (certain it's by a computer) to 10 (certain it's by a human)?''. Lastly, we asked participants for a yes/no response to the question, ``Are there any clearly awkward or weird portions of the analysis?'' asking if they would optionally explain their response if any were found. Results are in Table \ref{table:gpa}.

\subsubsection{Ablation Experiments and Case Study}

For ablation experiments (see Appendix A), we removed certain node and edge features, then trained 30 model copies for 3 epochs, evaluating maximum validation accuracy for each ablated model over the combined 90 epochs the model copies were trained. We additionally explored how model performance changes with different values of threshold $c_{\min}$ and edge direction weighting $\alpha$. Our case study (see Appendix B) shows a comparison of \ourmethod{}'s analysis of Pachelbel's \textit{Primi Toni 1} with that of the human analyst.

\subsection{Results} 

\subsubsection{Relative Performance} Table \ref{table:baselines_vs_model} presents performance of AutoSchA vs. various existing models, suggesting higher best-case accuracy and low best-case monotonicity loss for our proposed methodology. Notably, our model with sequential encoding for global features attains the highest accuracy.


\begin{table}[h!]
\centering
\begin{tabular}{c | c c}
Method & Accuracy & Monotonicity \\ 
\hline\hline
MLP & 0.714 (0.002) & 0.014 (0.001) \\ 
Transformer & 0.707 (0.011) & 0.011 (0.004) \\
GCN &  0.663 (0.006) & 0.007 (0.002) \\
GAT & 0.675 (0.006) &  0.012 (0.018)\\
R-GCN & 0.729 (0.012) & 0.013 (0.002) \\
\hline
 \ourmethod{} (Base) & 0.731 (0.015) & \textbf{0.003 (0.002)} \\
 \ourmethod{} (Grassmann) & \textit{0.738 (0.021)} & 0.016 (0.020) \\
\ourmethod{} (Sequence) & \textbf{0.749 (0.015)} & \textit{0.004 (0.003)} 
\end{tabular}
\caption{Mean (SD) best-case validation set accuracy (higher is better) and monotonicity loss (lower is better).}
\label{table:baselines_vs_model}
\vspace{-2mm}
\end{table}

\subsubsection{Human Experiments} Table \ref{table:gpa} suggests that \ourmethod{} is significantly better than the flawed analysis and relatively close to the performance of the human analyst. The average grade for \ourmethod{} is around a B+, whereas the human received around an A-, and the flawed analysis a C. Similarly, when judging musicality, \ourmethod{} again nears the level of the human analyst, easily outperforming the flawed model. Although participants tended to believe most analyses were human-generated, \textit{\ourmethod{} was rated similarly to human analyses} while the flawed analyses were statistically dissimilar. Finally, regarding the presence of ``weird analyses,'' \ourmethod{} statistically outperforms the flawed analysis, but still has a significantly larger number of perceived awkward components compared with the human analyses, suggesting room for future work. 

\begin{table}[h!]
\vspace{-2mm}
\centering
\begin{tabular}{l c c r} 
Method & Mean & 95\% CI & p-value \\ 
 \hline\hline
 \multicolumn{4}{c}{Grade Point Average (higher is better)}\\
 \ourmethod{} & 3.216 & (3.032, 3.399) & ref. \\ 
 Human & 3.551 & (3.363, 3.740) & 0.012 \\
 Flawed & 2.158 & (1.871, 2.445) & \textless 0.001 \\
 \hline
 \multicolumn{4}{c}{Musicality (higher is better)}\\
  \ourmethod{} & 7.196 & (6.682, 7.709) & ref. \\ 
 Human & 8.038 & (7.469, 8.607) & 0.029 \\
 Flawed & 4.653 & (4.022, 5.284) & \textless 0.001 \\
 \hline
 \multicolumn{4}{c}{Confidence in Human Annotation (higher is better)}\\
  Human & 6.467 & (5.824, 7.109) & ref. \\ 
 \ourmethod{} & 6.227 & (5.654, 6.800) & 0.672 \\
 Flawed & 5.236 & (4.597, 5.874) & 0.010 \\
 \hline
 \multicolumn{4}{c}{``Weird'' Analysis Proportion (lower is better)}\\
  \ourmethod{} & 0.667 & (0.523, 0.810) & ref. \\ 
 Human & 0.333 & (0.190, 0.477) & 0.001 \\
 Flawed & 0.956 & (0.893, 1.) & \textless 0.001 \\
 \end{tabular}
\caption{Evaluation of AutoSchA performance metrics.}
\label{table:gpa}
\vspace{-4mm}
\end{table}

\subsubsection{Ablation Experiments} Our ablation experiments suggest that rhythmic features are most vital to model performance, and including too many pitch features seems to confuse the model. Maximum accuracy was achieved by removing two of the three pitch features: pitch class and midi. We hypothesize that larger, more complex musical scores, and scores of different styles, may show very different variable importance metrics. For instance, a style that includes many overlapping suspensions would not rely on metric strength as much, for important structural tones would often be offset from strong metrical beats. Regarding threshold and edge direction weighting, the ideal threshold is 0.5 and the optimal alpha is 0.75, indicating that music analysis is mostly a forwards-looking phenomenon. The backwards component is also essential based on the steep drop from alpha=0.75 to alpha=1.0.

\section{Conclusion}
\label{sec:conclusion}

In this study, we describe a novel deep learning framework capable of generating convincing Schenkerian analyses near the level of a human analyst. These can be improved by human analysts more easily than creating them from scratch, and they can be used for downstream tasks like music generation and a broad set of applications to music theory analysis. We build on recent developments in GNNs and hierarchical deep learning to represent hierarchical musical analysis as a graph pooling problem, along with a novel graph pooling mechanism to directly model and learn the pooling process. By demonstrating the benefits of graph data structures for automatic hierarchical music representation, we open exciting new research avenues that advance music theory and methods development via GNNs.

\bibliography{aaai2026}

\end{document}